\documentclass[a4paper]{article}
\usepackage{setspace}
\usepackage[margin=1in]{geometry}
\usepackage{amsmath,amsfonts,graphicx,enumitem, booktabs, tikz, xcolor}
\usepackage[ruled,linesnumbered]{algorithm2e}
\usepackage[sectionbib]{natbib}
\usepackage[para]{threeparttable}

\allowdisplaybreaks[1]
\definecolor{mypurple}{RGB}{160, 32, 240}

\providecommand{\keywords}[1]
{
  \small	
  \textbf{\textit{Keywords:}} #1
}

\begin{document}

\title{Scalable likelihood-based estimation and variable selection for the Cox model with incomplete covariates}

\author{Ngok Sang Kwok and Kin Yau Wong\thanks{Corresponding author; email: kin-yau.wong@polyu.edu.hk}\\[4pt]Department of Applied Mathematics, The Hong Kong Polytechnic University}

\date{}

\maketitle

\begin{abstract}
Regression analysis with missing data is a long-standing and challenging problem, particularly when there are many missing variables with arbitrary missing patterns. Likelihood-based methods, although theoretically appealing, are often computationally inefficient or even infeasible when dealing with a large number of missing variables. In this paper, we consider the Cox regression model with incomplete covariates that are missing at random. We develop an expectation-maximization (EM) algorithm for nonparametric maximum likelihood estimation, employing a transformation technique in the E-step so that it involves only a one-dimensional integration. This innovation makes our methods scalable with respect to the dimension of the missing variables. In addition, for variable selection, we extend the proposed EM algorithm to accommodate a LASSO penalty in the likelihood. We demonstrate the feasibility and advantages of the proposed methods over existing methods by large-scale simulation studies and apply the proposed methods to a cancer genomic study.
\end{abstract}


\section{Introduction}
\label{sec:intro}


In many public health and medical studies, we study the association between covariates, such as treatment received, demographic information, or other personal characteristics, and times to disease events or death. One complication that often arises in practice is that the covariates may not be available for all study subjects. For example, The Cancer Genome Altas (TCGA) collected genomic and clinical data for many types of cancer, but for a substantial number of subjects, protein expressions were not measured. Another example is the North Staffordshire Osteoarthritis Project (\citeauthor{wilkie2019reasons}, \citeyear{wilkie2019reasons}), where researchers investigated the association between mortality and symptomatic osteoarthritis. Covariates such as walking frequency, depression, and BMI, collected via questionnaires, had missing responses from some subjects. Missing data pose significant theoretical and computational challenges for regression analysis.

There are multiple popular approaches for handling missing covariates in regression analyses, many of which have been applied to survival analysis. One is the likelihood approach, where we incorporate a model for the missing covariates into the likelihood. An advantage of this approach is that maximum likelihood estimation (MLE) is efficient. 
\cite{herring2001likelihood} considered the Cox model with incomplete covariates, which could be categorical or continuous, and developed an expectation-maximization (EM) algorithm (\citeauthor{dempster1977maximum}, \citeyear{dempster1977maximum}) for computation.
\cite{zhou2022new} implemented the EM algorithm with two-stage data augmentation for the Cox model with interval-censored survival time. However, MLE becomes computationally intensive or even infeasible with a large number of missing covariates. 

Another approach is inverse-probability weighting (\citeauthor{martinussen2016cox}, \citeyear{martinussen2016cox}), where only subjects with complete observations are used and are weighted according to the probability of complete observation. \cite{thiessen2022unified} proposed a two-step estimator for the Cox model with MAR incomplete covariates, where two inverse-probability weighted estimators are combined to form a more efficient estimator. Inverse probability weighting, however, is in general inefficient as it discards information contained in subjects with partial observations.

A third approach is (multiple) imputation, where the missing values are first imputed based on the observed data, and standard analysis methods can be applied to the completed data (\citeauthor{MultipleImputation}, \citeyear{MultipleImputation}). \cite{white2009imputing} considered multiple imputation with the Cox model. A popular version of multiple imputation is multiple imputation by chained equation (MICE), also known as a ``fully conditional specification,'' which involves specifying the conditional density for each incomplete covariate (\citeauthor{van2011mice}, \citeyear{van2011mice}; \citeauthor{raghunathan2001multivariate}, \citeyear{raghunathan2001multivariate}; \citeauthor{azur2011multiple}, \citeyear{azur2011multiple}). Another imputation technique is matrix-completion, which imputes the missing values through solving a rank restricted optimization problem (\citeauthor{hastie2015matrix}, \citeyear{hastie2015matrix}). However, the theoretical properties of estimators based on imputed data are not guaranteed.

In addition to missing data, high-dimensional covariates present another challenge. When the number of available covariates is large, standard approaches that regress on all covariates, such as MLE or estimating equations based on inverse-probability weighting, may suffer from over-fitting and difficulties in interpretations, or may even be infeasible. In such cases, we are often interested in selecting a subset of covariates that are associated with the outcome. Penalized regression methods, such as LASSO (\citeauthor{tibshirani1996regression}, \citeyear{tibshirani1996regression}), are popular approaches to reduce over-fitting and to perform variable selection. 


Penalized regression or variable selection with missing data is highly challenging, with limited research in this area. For likelihood-based methods, \cite{garcia2010variable} developed EM algorithms for the Cox model with LASSO, adaptive LASSO (\citeauthor{zou2006adaptive}, \citeyear{zou2006adaptive}), and the smoothly clipped absolute deviation (\citeauthor{fan2001variable}, \citeyear{fan2001variable}) penalties. \cite{sabbe2013emlasso} studied variable selection in logistic regression using a LASSO penalty via a stochastic EM algorithm.
For inverse probability weighting, \cite{johnson2008penalized} and \cite{wolfson2011eeboost} incorporated a penalty term to inverse-probability-weighted estimating equations for performing variable selection. For multiple imputation, \cite{wood2008should} investigated methods for combining variable selection results from multiply imputed datasets. \cite{deng2016multiple} extended the MICE approach to high-dimensional settings by fitting a penalized regression model for each missing covariate. \cite{liang2024variable} developed an iterative imputation method based on matrix completion and a randomized LASSO method based on bootstrap. However, these approaches suffer the shortcomings of their unpenalized counterparts, such as computational or estimation inefficiency and the lack of theoretical justifications. Also, they may require computationally intensive tuning.

In this paper, we study the likelihood approach and develop a novel algorithm for MLE that overcomes the computational challenges of existing methods. In particular, we consider the Cox proportional hazards model with incomplete covariates and develop an EM algorithm for computation, which is computationally feasible even when the missing pattern is arbitrary and a large number of covariates are missing for a subject. The algorithm involves a subject-specific transformation of the missing covariates, which results in a one-dimensional numerical integration in the E-step.
This is a major advance over existing likelihood-based methods, which involve numerical integration of the same dimension as the covariates and thus are feasible only for a small number of missing covariates.

Under the likelihood framework, we can naturally perform variable selection by incorporating a penalty term. In particular, we consider a LASSO penalty and develop an EM algorithm for computation. Employing the transformation technique, the E-step involves only a one-dimensional numerical integration, as in the unpenalized case. In the M-step, we perform quadratic approximation of the log-partial likelihood and adopt the coordinate-descent algorithm. Consequently, the estimation and variable selection procedure remain computationally feasible even with many missing covariates.

This paper is structured as follows. In Section 2, we describe the proposed model and formulate the EM algorithm for both the unpenalized and penalized cases. In Section 3, we perform large-scale simulation studies and compare the performance of the proposed methods with existing methods. In Section 4, we demonstrate the feasibility and advantages of the proposed method on a cancer genomics study. We provide some concluding remarks and possible extensions in Section 5.

\section{Methods}
\label{sec:method}
\subsection{Model and likelihood}
Let $T$ be an event time of interest and $\boldsymbol{X}$ be a $p$-vector of covariates. Assume that $T$ given $\boldsymbol{X}$ follows the Cox proportional hazards model, with the hazard function given by
\[
\lambda(t\mid\boldsymbol{X})=\lambda(t)e^{\boldsymbol{X}^{\mathrm{T}}\boldsymbol{\beta}},
\]
where $\boldsymbol{\beta}$ is a vector of regression parameters, and $\lambda(\cdot)$ is a nonparametric baseline hazard function. We assume that $\boldsymbol{X}$ follows the multivariate normal distribution with mean $\boldsymbol{\mu}$ and covariance matrix $\boldsymbol{\Sigma}$. Here, we for simplicity of presentation impose a parametric model on all components of $\boldsymbol{X}$, but the proposed methods can be easily generalized to the milder condition that a subset of components of $\boldsymbol{X}$, $\boldsymbol{X}_\mathcal{S}$, is multivariate normal given the remaining components, $\boldsymbol{X}_{-\mathcal{S}}$, if $\boldsymbol{X}_{-\mathcal{S}}$ is always observed.
Suppose that $T$ may be subject to right-censoring. Let $C$ be the censoring time, $Y=\min(T,C)$, and $\Delta=I(T\le C)$.

We allow components of $\boldsymbol{X}$ to be missing. Let $\boldsymbol{R}\equiv(R_1,\ldots,R_p)^{\mathrm{T}}$ denote a vector of missing indicators, where $R_j=1$ if $X_j$ is missing and $R_j=0$ if otherwise. Assume missing at random, such that $\boldsymbol{R}$ and $\boldsymbol{X}$ are independent given $\{X_j:P(R_j=0)=1\}$, $Y$, and $\Delta$. Also, assume that $T$ and $C$ are independent given $\{X_j:P(R_j=0)=1\}$. For a sample of size $n$, the observed data consist of $\mathcal{O}_i\equiv\{Y_i,\Delta_i,\boldsymbol{R}_i,\boldsymbol{X}_{i,-\boldsymbol{R}_i}\}$ for $i=1,\ldots,n$, where $\boldsymbol{X}_{i,-\boldsymbol{R}_i}$ denote the subvector of $\boldsymbol{X}_i$ consisting of components that correspond to $R_{ij}=0$. Let $\Lambda(t)=\int_0^t\lambda(s)\,\mathrm{d}s$ and $\boldsymbol{\theta}\equiv(\boldsymbol{\beta},\Lambda,\boldsymbol{\mu},\boldsymbol{\Sigma})$ denote the set of all unknown parameters. The (observed-data) likelihood is
\[
L{_\mathrm{obs}}(\boldsymbol{\theta})=\prod_{i=1}^n\int\big\{\lambda(Y_i)e^{\boldsymbol{X}_i^{\mathrm{T}}\boldsymbol{\beta}}\big\}^{\Delta_i}e^{-\Lambda(Y_i)e^{\boldsymbol{X}_i^{\mathrm{T}}\boldsymbol{\beta}}}|\boldsymbol{\Sigma}|^{-1/2}e^{-\frac{1}{2}(\boldsymbol{X}_i-\boldsymbol{\mu})^{\mathrm{T}}\boldsymbol{\Sigma}^{-1}(\boldsymbol{X}_i-\boldsymbol{\mu})}\,\mathrm{d}\boldsymbol{X}_{i,\boldsymbol{R}_i},
\]
where $\boldsymbol{X}_{i,\boldsymbol{R}_i}$ denote the subvector of $\boldsymbol{X}_i$ consisting of the components that correspond to $R_{ij}=1$.

We adopt the nonparametric likelihood estimation (NPMLE) approach. Let $t_1<\cdots<t_m$ be the ordered unique observed event times, where $m=\sum_{i=1}^n\Delta_i$. We set $\Lambda$ to be a step function that jumps only at $t_1,\ldots,t_m$ and let the corresponding jump sizes be $\lambda_1,\ldots,\lambda_m$. In the likelihood, we replace $\lambda(Y_i)$ by the corresponding jump size. In the sequel, we use $L{_\mathrm{obs}}$ to denote this nonparametric version of the likelihood.

\subsection{EM algorithm for unpenalized estimation}
When the dimension of $\boldsymbol{X}$ is low and we are not interested in variable selection, we estimate $\boldsymbol{\theta}$ by the NPMLE $(\widehat{\boldsymbol{\beta}},\widehat{\Lambda},\widehat{\boldsymbol{\mu}},\widehat{\boldsymbol{\Sigma}})$, which is the maximizer of $L_{\mathrm{obs}}$.
We adopt the EM algorithm to compute the NPMLE, with $\boldsymbol{X}_{i,\boldsymbol{R}_i}$ treated as missing data for $i=1,\ldots,n$. The complete-data log-likelihood is
\[
\log L_{\mathrm{com}}(\boldsymbol{\theta})=\sum_{i=1}^n\Big\{\Delta_i(\log\lambda_{j(i)}+\boldsymbol{X}_i^{\mathrm{T}}\boldsymbol{\beta})-\sum_{j:t_j\le Y_i}\lambda_je^{\boldsymbol{X}_i^{\mathrm{T}}\boldsymbol{\beta}}-\frac{1}{2}\log|\boldsymbol{\Sigma}|-\frac{1}{2}(\boldsymbol{X}_i-\boldsymbol{\mu})^{\mathrm{T}}\boldsymbol{\Sigma}^{-1}(\boldsymbol{X}_i-\boldsymbol{\mu})\Big\},
\]
where $j(i)$ is such that $Y_{j(i)}=t_j$ for $i\in\{i=1,\ldots,n:\Delta_i=1\}$.
In the E-step, we evaluate the conditional expectation of $\log L_{\mathrm{com}}(\boldsymbol{\theta})$ given the observed data at the current parameter estimate. In the M-step, we maximize the expected complete-data log-likelihood. In particular, at the $(k+1)$th iteration, we update
\begin{align}
\boldsymbol{\mu}^{(k+1)} &\,= \frac{1}{n} \sum_{i=1}^n \widehat{\mathrm{E}}^{(k)}(\boldsymbol{X}_i)\label{eq:normalmu}\\
\boldsymbol{\Sigma}^{(k+1)} &\,= \frac{1}{n} \sum_{i=1}^n\widehat{\mathrm{E}}^{(k)}(\boldsymbol{X}_i\boldsymbol{X}_i^{\mathrm{T}}) - \left(\boldsymbol{\mu}^{(k+1)}\right) \left(\boldsymbol{\mu}^{(k+1)}\right)^{\mathrm{T}}\label{eq:normalsigma},
\end{align}
where $\widehat{\mathrm{E}}^{(k)}$ denote conditional expectation given the observed data, evaluated at the parameter estimate at the $k$th iteration.
After profiling out $\lambda_1,\ldots,\lambda_m$, $\boldsymbol{\beta}$ maximizes the following ``complete-data log-partial likelihood''
\[
Q^{(k)}(\boldsymbol{\beta}) = \sum_{i=1}^n \Delta_i\left[\widehat{\mathrm{E}}^{(k)}(\boldsymbol{X}_i)^{\mathrm{T}}\boldsymbol{\beta}-\log\bigg\{ \sum_{j:Y_j\le Y_i} \widehat{\mathrm{E}}^{(k)}(e^{\boldsymbol{X}_j^{\mathrm{T}}\boldsymbol{\beta}})\bigg\} \right].
\]
Note that
\begin{align*}
\frac{\partial Q^{(k)}(\boldsymbol{\beta})}{\partial \boldsymbol{\beta}} =&\, \sum_{i=1}^n \Delta_i\left\{\widehat{\mathrm{E}}^{(k)}(\boldsymbol{X}_i) - \frac{\sum_{j:Y_j\le Y_i}\widehat{\mathrm{E}}^{(k)}(e^{\boldsymbol{X}_j^{\mathrm{T}}\boldsymbol{\beta}}\boldsymbol{X}_j)}{\sum_{j:Y_j\le Y_i}\widehat{\mathrm{E}}^{(k)}(e^{\boldsymbol{X}_j^{\mathrm{T}}\boldsymbol{\beta}})} \right\}\\
\frac{\partial^2 Q^{(k)}(\boldsymbol{\beta})}{\partial \boldsymbol{\beta}\partial\boldsymbol{\beta}^{\mathrm{T}}} =&\, - \sum_{i=1}^n \Delta_i\Bigg[
\frac{\sum_{j:Y_j\le Y_i}  \widehat{\mathrm{E}}^{(k)}(e^{\boldsymbol{X}_j^{\mathrm{T}}\boldsymbol{\beta}}\boldsymbol{X}_j\boldsymbol{X}_j^{\mathrm{T}})}{\sum_{j:Y_j\le Y_i} \widehat{\mathrm{E}}^{(k)}( e^{\boldsymbol{X}_j^{\mathrm{T}}\boldsymbol{\beta}})} 
-\Bigg\{\frac{\sum_{j:Y_j\le Y_i} \widehat{\mathrm{E}}^{(k)}(e^{\boldsymbol{X}_j^{\mathrm{T}}\boldsymbol{\beta}}\boldsymbol{X}_j)}{\sum_{j:Y_j\le Y_i} \widehat{\mathrm{E}}^{(k)}(e^{\boldsymbol{X}_j^{\mathrm{T}}\boldsymbol{\beta}})}\Bigg\}^{\otimes2}\Bigg].
\end{align*}
We update $\boldsymbol{\beta}$ by the one-step Newton method:
\begin{equation}\label{eq:newton}
\boldsymbol{\beta}^{(k+1)} = \boldsymbol{\beta}^{(k)} - \left( \frac{\partial^2 Q^{(k)}(\boldsymbol{\beta})}{\partial\boldsymbol{\beta}\partial\boldsymbol{\beta}^{\mathrm{T}}} \bigg\vert_{\boldsymbol{\beta}=\boldsymbol{\beta}^{(k)}} \right)^{-1}\left( \frac{\partial Q^{(k)}(\boldsymbol{\beta})}{\partial\boldsymbol{\beta}} \bigg\vert_{\boldsymbol{\beta}=\boldsymbol{\beta}^{(k)}}\right),
\end{equation}
where $\boldsymbol{\beta}^{(k)}$ denote the estimate of $\boldsymbol{\beta}$ at the $k$th iteration.
Finally, we update the baseline hazard function using the Breslow-like estimator:
\begin{equation}\label{eq:lambda}
\lambda_{j(i)}^{(k+1)} = \frac{1}{\sum_{j:Y_j\le Y_i} \widehat{\mathrm{E}}^{(k)}(e^{\boldsymbol{X}_j^{\mathrm{T}} \boldsymbol{\beta}^{(k+1)}})} 
\end{equation}
for $i$ such that $\Delta_i=1$.

The major computational challenge of the EM algorithm is that the conditional distribution of $\boldsymbol{X}_i$ does not have a closed form, and direct numerical integration for the expectations is infeasible when the dimension of $\boldsymbol{X}_{i,\boldsymbol{R}_i}$ is moderately high. To avoid multi-dimensional numerical integrations, we propose a transformation approach under which the expectations can be computed using at most one-dimensional numerical integrations. Let $\boldsymbol{\beta}_{\boldsymbol{R}_i}$ and $\boldsymbol{\beta}_{-\boldsymbol{R}_i}$ denote the subvector of $\boldsymbol{\beta}$ consisting of components that correspond to $R_{ij}=1$ and 0 respectively. The same method is used to denote subvectors of $\boldsymbol{\mu}$. Let $\boldsymbol{\Sigma}_{\mathcal{A}_{i1}, \mathcal{A}_{i2}}$ denote the submatrix of $\boldsymbol{\Sigma}$ with rows indexed by $\mathcal{A}_{i1}$ and columns indexed by $\mathcal{A}_{i2}$, and $\mathcal{A}_{ij}$ is either $\boldsymbol{R}_i$ or $-\boldsymbol{R}_i$ for $j=1,2$. The expectations that need to be computed in the E-step are in one of the following forms:
\begin{equation}\label{eq:1}
\mathrm{E}\big\{\exp(\boldsymbol{X}_{i,\boldsymbol{R}_i}^{\mathrm{T}} \boldsymbol{\beta}^{(k)}_{\boldsymbol{R}_i})\vert \mathcal{O}_i\big\}
\end{equation}
\begin{equation}\label{eq:2}
\mathrm{E}\{g(\boldsymbol{X}_{i, \boldsymbol{R}_i}^{\mathrm{T}} \boldsymbol{\beta}^{(k)}_{\boldsymbol{R}_i})\boldsymbol{X}_{i, \boldsymbol{R}_i} \vert \mathcal{O}_i\}
\end{equation}
\begin{equation}\label{eq:3}
\mathrm{E}\{g(\boldsymbol{X}_{i, \boldsymbol{R}_i}^{\mathrm{T}} \boldsymbol{\beta}^{(k)}_{\boldsymbol{R}_i})\boldsymbol{X}_{i,\boldsymbol{R}_i}\boldsymbol{X}_{i, \boldsymbol{R}_i}^{\mathrm{T}}\vert \mathcal{O}_i\}
\end{equation}
\begin{equation}\label{eq:4}
\mathrm{E}\big\{\exp(\boldsymbol{X}_{i,\boldsymbol{R}_i}^{\mathrm{T}} \boldsymbol{\beta}^{(k+1)}_{\boldsymbol{R}_i})\vert \mathcal{O}_i\big\},
\end{equation}
where $g$ is either the exponential function or the constant function $g(\cdot)=1$. Note that the expectations are evaluated at $\boldsymbol{\theta}=\boldsymbol{\theta}^{(k)}$.

First, if $\boldsymbol{\beta}^{(k)}_{\boldsymbol{R}_i}=\boldsymbol{0}$, then the conditional distribution of $\boldsymbol{X}_{i,\boldsymbol{R}_i}$ given $\mathcal{O}_i$ is a multivariate normal distribution that does not depend on $(Y_i,\Delta_i)$. The expectations (\ref{eq:1})--(\ref{eq:4}) have simple closed-form expressions.

For $\boldsymbol{\beta}^{(k)}_{\boldsymbol{R}_i} \neq \boldsymbol{0}$, we define an orthogonal matrix $\boldsymbol{\Psi}_i$ with the first row being $ (\boldsymbol{\beta}^{(k)}_{\boldsymbol{R}_i})^{\mathrm{T}}/ \Vert\boldsymbol{\beta}^{(k)}_{\boldsymbol{R}_i}\Vert$ and let $\widetilde{\boldsymbol{X}}_i = \boldsymbol{\Psi}_i \boldsymbol{X}_{i, \boldsymbol{R}_i}$, where $\Vert\cdot\Vert$ denote the $L_2$-norm. Note that the first component of $\widetilde{\boldsymbol{X}}_i$ is $\widetilde{X}_{i1} = \boldsymbol{X}_{i, \boldsymbol{R}_i}^\mathrm{T} \boldsymbol{\beta}^{(k)}_{\boldsymbol{R}_i} / \Vert\boldsymbol{\beta}^{(k)}_{\boldsymbol{R}_i}\Vert$. Let $\boldsymbol{\eta}_i$, and $\boldsymbol{\nu}_i$ denote the mean and variance of $\widetilde{\boldsymbol{X}}_i$ given $\boldsymbol{X}_{-\boldsymbol{R}_i}$, where
\begin{align*}
    \boldsymbol{\eta}_i &= \boldsymbol{\Psi}_i\boldsymbol{\mu}_{\boldsymbol{R}_i} + \boldsymbol{\Psi}_i\boldsymbol{\Sigma}_{\boldsymbol{R}_i, -\boldsymbol{R}_i}\boldsymbol{\Sigma}_{-\boldsymbol{R}_i, -\boldsymbol{R}_i}^{-1} (\boldsymbol{X}_{i,-\boldsymbol{R}_i} - \boldsymbol{\mu}_{-\boldsymbol{R}_i})\\
    \boldsymbol{\nu}_i &=  \boldsymbol{\Psi}_i\boldsymbol{\Sigma}_{\boldsymbol{R}_i, \boldsymbol{R}_i}\boldsymbol{\Psi}_i^\mathrm{T} - \boldsymbol{\Psi}_i\boldsymbol{\Sigma}_{\boldsymbol{R}_i, -\boldsymbol{R}_i}\boldsymbol{\Sigma}_{-\boldsymbol{R}_i, -\boldsymbol{R}_i}^{-1} \boldsymbol{\Sigma}_{-\boldsymbol{R}_i, \boldsymbol{R}_i}\boldsymbol{\Psi}_i^\mathrm{T}.
\end{align*}
Let $\widetilde{\boldsymbol{X}}_{i,-1}$ denote the subvector of $\widetilde{\boldsymbol{X}}_i$ consisting of all but the first component.
Although the conditional distribution of $\widetilde{\boldsymbol{X}}_i$ given the observed data does not have a simple form, $\widetilde{\boldsymbol{X}}_{i,-1}$ given the observed data and $\widetilde{X}_{i1}$ (at $\boldsymbol{\theta}=\boldsymbol{\theta}^{(k)}$) follows the multivariate normal distribution:
\begin{align*}
\widetilde{\boldsymbol{X}}_{i,-1}\mid(Y_i,\Delta_i,\widetilde{X}_{i1},\boldsymbol{X}_{i,-\boldsymbol{R}_i})\sim &\,\mathrm{N}\Bigg( (\boldsymbol{\eta}_i)_{-1} + (\boldsymbol{\nu}_i)_{-1,1}\frac{\widetilde{X}_{i1}-(\boldsymbol{\eta}_i)_{1}}{(\boldsymbol{\nu}_i)_{1,1}},  (\boldsymbol{\nu}_i)_{-1,-1}-\frac{(\boldsymbol{\nu}_i)_{-1,1}^{\otimes2}}{(\boldsymbol{\nu}_i)_{1,1}}\Bigg)\\
\equiv&\, \mathrm{N}(\boldsymbol{m}_i(\widetilde{X}_{i1}),\boldsymbol{V}_i),
\end{align*}
where $(\boldsymbol{\nu}_i)_{1,1}$ is the upper left element of $\boldsymbol{\nu}_i$, $(\boldsymbol{\nu}_i)_{-1,1}$ is the first column of $\boldsymbol{\nu}_i$ with the first component removed, and $(\boldsymbol{\nu}_i)_{-1,-1}$ is the lower right submatrix of $\boldsymbol{\nu}_i$, with the first row and column of $\boldsymbol{\nu}_i$ removed.
The conditional density of $\widetilde{X}_{i1}$ given $\mathcal{O}_i$ is proportional to
\[
f(\widetilde{x}_{i1}; \mathcal{O}_i)\equiv\exp\left\{\Delta_i\Vert\boldsymbol{\beta}^{(k)}_{\boldsymbol{R}_i}\Vert\widetilde{x}_{i1} - \Lambda_0(Y_i)e^{\Vert\boldsymbol{\beta}^{(k)}_{\boldsymbol{R}_i}\Vert\widetilde{x}_{i1} +  \boldsymbol{X}_{i, -\boldsymbol{R}_i}^{\mathrm{T}} \boldsymbol{\beta}_{-\boldsymbol{R}_i}} - \frac{1}{2 (\boldsymbol{\nu}_i)_{1,1}}(\widetilde{x}_{i1}- (\boldsymbol{\eta}_i)_1)^2  \right\}.
\]
Therefore, conditional expectations of functions of $\boldsymbol{X}_{i,\boldsymbol{R}_i}\equiv \boldsymbol{\Psi}_i^{\mathrm{T}}\widetilde{\boldsymbol{X}}_i$ can be computed by first further conditioning on $\widetilde{X}_{i1}$, where the conditional expectations have closed-form expressions, and then taking the expectation over $\widetilde{X}_{i1}$, which can be performed by numerical integration.

Specifically, the expectation (\ref{eq:1}) is equal to $\mathrm{E}\big\{\exp(\Vert\boldsymbol{\beta}^{(k)}_{\boldsymbol{R}_i}\Vert\widetilde{X}_{i1})\vert \mathcal{O}_i\big\}$.
The expectation (\ref{eq:2}) is equal to
\[
\boldsymbol{\Psi}_i^{\mathrm{T}}\mathrm{E}\big\{g(\Vert\boldsymbol{\beta}^{(k)}_{\boldsymbol{R}_i}\Vert\widetilde{X}_{i1})\widetilde{\boldsymbol{X}}_i\vert \mathcal{O}_i\big\}=\boldsymbol{\Psi}_{i}^{\mathrm{T}}\mathrm{E}\left\{ g(\Vert\boldsymbol{\beta}_{\boldsymbol{R}_{i}}^{(k)}\Vert\widetilde{X}_{i1})\left(\begin{array}{c}
\widetilde{X}_{i1}\\
\boldsymbol{m}_{i}(\widetilde{X}_{i1})
\end{array}\right)\Bigg|\mathcal{O}_{i}\right\}.
\]
The expectation (\ref{eq:3}) is equal to
\begin{align*}\label{eq:7}
&\,\boldsymbol{\Psi}_i^{\mathrm{T}}\mathrm{E}\big\{g(\Vert\boldsymbol{\beta}^{(k)}_{\boldsymbol{R}_i}\Vert\widetilde{X}_{i1})\widetilde{\boldsymbol{X}}_i\widetilde{\boldsymbol{X}}_i^{\mathrm{T}}\vert \mathcal{O}_i\big\}\boldsymbol{\Psi}_i\\
=&\,\boldsymbol{\Psi}_{i}^{\mathrm{T}}\mathrm{E}\left\{ g(\Vert\boldsymbol{\beta}_{\boldsymbol{R}_{i}}^{(k)}\Vert\widetilde{X}_{i1})\left(\begin{array}{cc}
\widetilde{X}_{i1}^{2} & \widetilde{X}_{i1}\boldsymbol{m}_{i}(\widetilde{X}_{i1})^{\mathrm{T}}\\
\widetilde{X}_{i1}\boldsymbol{m}_{i}(\widetilde{X}_{i1}) & \boldsymbol{V}_{i}+\boldsymbol{m}_{i}(\widetilde{X}_{i1})\boldsymbol{m}_{i}(\widetilde{X}_{i1})^{\mathrm{T}}
\end{array}\right)\Bigg|\mathcal{O}_{i}\right\} \boldsymbol{\Psi}_{i}.
\end{align*}
Finally, to evaluate (\ref{eq:4}), let
\[
\phi_i(\widetilde{X}_{i1};\boldsymbol{a})=\mathrm{E}\big\{\exp(\widetilde{\boldsymbol{X}}_{i,-1}^{\mathrm{T}}\boldsymbol{a})\mid \widetilde{X}_{i1},\mathcal{O}_i\big\}=\exp\Big\{\boldsymbol{a}^{\mathrm{T}}\boldsymbol{m}_i(\widetilde{X}_{i1})+\frac{1}{2}\boldsymbol{a}^{\mathrm{T}}\boldsymbol{V}_i\boldsymbol{a}\Big\}
\]
for any vector $\boldsymbol{a}$ of an appropriate dimension. We can write (\ref{eq:4}) as
\[
\mathrm{E}\big\{\exp(\widetilde{\boldsymbol{X}}_i^{\mathrm{T}} \boldsymbol{\Psi}_i\boldsymbol{\beta}^{(k+1)}_{\boldsymbol{R}_i})\vert \mathcal{O}_i\big\}=\mathrm{E}\big\{\exp((\boldsymbol{\Psi}_i\boldsymbol{\beta}^{(k+1)}_{\boldsymbol{R}_i})_1\widetilde{X}_{i1})\phi_i(\widetilde{X}_{i1};(\boldsymbol{\Psi}_i\boldsymbol{\beta}^{(k+1)}_{\boldsymbol{R}_i})_{-1})\mid\mathcal{O}_i\big\}.
\]
Therefore, all expectations involved in the E-step can be computed using one-dimensional numerical integrations over the conditional distribution of $\widetilde{X}_{i1}$.
In particular, for any function $h$, we have
\[
\mathrm{E}\{ h(\widetilde{X}_{i1})\mid \mathcal{O}_i\} = \frac{\int h(\widetilde{x}_{i1}) f(\widetilde{x}_{i1};\mathcal{O}_i) \,\mathrm{d}\widetilde{x}_{i1} }{\int  f(\widetilde{x}_{i1};\mathcal{O}_i) \,\mathrm{d}\widetilde{x}_{i1} }.
\]
The integrations can be approximated using the adaptive Gauss--Hermite quadrature (\citeauthor{liu1994note}, \citeyear{liu1994note}). The proposed algorithm is summarized in Algorithm 1.

\begin{algorithm}
\caption{NPMLE}\label{NPMLE}
\DontPrintSemicolon
\SetKwInOut{Input}{Input}\SetKwInOut{Output}{Output}
\Input{$\{\mathcal{O}_i\}_{i=1,2,\hdots, n}$.}
Initialize $(\boldsymbol{\beta}^{(0)}, \boldsymbol{\lambda}^{(0)}, \boldsymbol{\mu}^{(0)}, \boldsymbol{\Sigma}^{(0)})$.\;
Calculate (\ref{eq:1}),  (\ref{eq:2}), and (\ref{eq:3}) for $i=1,2,\hdots, n$ and in turn the gradient and Hessian of $Q^{(k)}(\boldsymbol{\beta})$.\;
Update $\boldsymbol{\mu}$ and $\boldsymbol{\Sigma}$ by (\ref{eq:normalmu}) and (\ref{eq:normalsigma}), respectively.\;
Update $\boldsymbol{\beta}$ by (\ref{eq:newton}).\;
Calculate (\ref{eq:4}) for $i=1,2,\hdots, n$.\;
Update $\Lambda$ by (\ref{eq:lambda}).\;
Repeat Steps 2--6 until convergence.\;
\Output{$(\widehat{\boldsymbol{\beta}},\widehat{\Lambda},\widehat{\boldsymbol{\mu}},\widehat{\boldsymbol{\Sigma}})$.}
\end{algorithm}

\subsection{EM algorithm for penalized estimation}
When the number of covariates is large, it is often desirable to select a subset of covariates that are associated with the survival time. We propose a penalization approach with the following penalized observed-data log-likelihood:
\begin{equation*}
p\ell(\boldsymbol{\theta})=\log L_{\mathrm{obs}}(\boldsymbol{\theta}) - n \gamma\Vert\boldsymbol{\beta}\Vert_1,
\end{equation*}
where $\gamma>0$ is tuning parameter. The penalized NPMLE is the maximizer of $p\ell(\boldsymbol{\theta})$. Note that we assume that the sample size is sufficiently larger than the number of covariates, so no penalty is imposed for the covariance matrix $\boldsymbol{\Sigma}$. 

To compute the penalized NPMLE, we adopt the proposed EM algorithm for the unpenalized estimator with some modifications. The E-step for the penalized estimator is the same as the previous algorithm. In the M-step, the estimators of $\boldsymbol{\mu}$ and $\boldsymbol{\Sigma}$ are the same as before.

After profiling out the baseline hazard function, $\boldsymbol{\beta}$ maximizes the (expected) penalized complete-data log-partial likelihood $n^{-1}Q^{(k)}(\boldsymbol{\beta})-\gamma\Vert\boldsymbol{\beta}\Vert_1$. Clearly, there is no closed-form solution, and the objective function is not differentiable. To update $\boldsymbol{\beta}$, we first approximate the objective function using a second-order Taylor expansion:
\begin{align}
n^{-1}Q^{(k)}(\boldsymbol{\beta}) - \gamma\Vert\boldsymbol{\beta}\Vert_1 \approx  -\frac{1}{2} \boldsymbol{\beta}^{\mathrm{T}} \boldsymbol{A} \boldsymbol{\beta} - \boldsymbol{P}^{\mathrm{T}} \boldsymbol{\beta} - \gamma\Vert\boldsymbol{\beta}\Vert_1 + \text{const},\label{eq:approx}
\end{align}
where 
\begin{align*}
\boldsymbol{A} =&\, -\frac{1}{n}\left(\frac{\partial^2 Q^{(k)}(\boldsymbol{\beta})}{\partial\boldsymbol{\beta}\partial\boldsymbol{\beta}^{\mathrm{T}}}\bigg\vert_{\boldsymbol{\beta}=\boldsymbol{\beta}^{(k)}}\right)\\
\boldsymbol{P} =&\, \frac{1}{n}\left\{\left(\frac{\partial^2 Q^{(k)}(\boldsymbol{\beta})}{\partial\boldsymbol{\beta}\partial\boldsymbol{\beta}^{\mathrm{T}}}\bigg\vert_{\boldsymbol{\beta}=\boldsymbol{\beta}^{(k)}}\right)\boldsymbol{\beta}^{(k)} - \left(\frac{\partial Q^{(k)}(\boldsymbol{\beta})}{\partial\boldsymbol{\beta}}\bigg\vert_{\boldsymbol{\beta}=\boldsymbol{\beta}^{(k)}}\right)\right\}.
\end{align*}
Then, to maximize the right-hand side of (\ref{eq:approx}), we adopt the coordinate-descent algorithm (\citeauthor{simon2011regularization}, \citeyear{simon2011regularization}). For $j=1,\ldots,p$, we update $\beta_j$ with $\boldsymbol{\beta}_{-j}$ fixed at the current estimates by setting
\begin{equation}\label{eq:coor}
\beta_j = -\frac{S(\boldsymbol{A}_{j,-j}\boldsymbol{\beta}_{-j}+\boldsymbol{P}_j, \gamma)}{\boldsymbol{A}_{j,j} },
\end{equation}
where $S(x, \gamma) = \text{sgn}(x)(\vert x\vert - \gamma)_+$. We iterate over components of $\boldsymbol{\beta}$ until convergence. After updating $\boldsymbol{\beta}$, we update $\Lambda$ using the same Breslow-like estimator as before. This completes a single M-step. Let $\boldsymbol{\beta}_{\gamma}$ be the LASSO estimator corresponding to $\gamma$ and $\mathfrak{B}_\gamma$ denote the active set of $\boldsymbol{\beta}_{\gamma}$. Since the LASSO estimator is biased, we refit the model using NPMLE, with only the coefficients in the active set $\mathfrak{B}_\gamma$ allowed to be nonzero. We summarize the procedure in Algorithm 2. 
\begin{algorithm}
\caption{Penalized  NPMLE}\label{Penalized NPMLE}
\DontPrintSemicolon
\SetKwInOut{Input}{Input}\SetKwInOut{Output}{Output}
\Input{$\{\mathcal{O}_i\}_{i=1,2,\hdots, n}$ and $\gamma$.}
Initialize $(\boldsymbol{\beta}^{(0)}, \boldsymbol{\lambda}^{(0)}, \boldsymbol{\mu}^{(0)}, \boldsymbol{\Sigma}^{(0)})$.\;
Calculate (\ref{eq:1}),  (\ref{eq:2}), and (\ref{eq:3}) for $i=1,2,\hdots, n$ and in turn the gradient and Hessian of $Q^{(k)}(\boldsymbol{\beta})$.\;
Update $\boldsymbol{\mu}$ and $\boldsymbol{\Sigma}$ by (\ref{eq:normalmu}) and (\ref{eq:normalsigma}), respectively.\;
Iteratively update each component of $\boldsymbol{\beta}$ through (\ref{eq:coor}) until convergence.\;
Calculate (\ref{eq:4}) for $i=1,2,\hdots, n$.\;
Update $\Lambda$ through (\ref{eq:lambda}).\;
Repeat Steps 2--6 until convergence.\;
Refit NPMLE over the active set.\;
\Output{$(\widehat{\boldsymbol{\beta}}_{\gamma}^\text{refit},\widehat{\Lambda}_{\gamma}^\text{refit},\widehat{\boldsymbol{\mu}}_{\gamma}^\text{refit},\widehat{\boldsymbol{\Sigma}}_{\gamma}^\text{refit})$.}
\end{algorithm}

We specify a grid of tuning parameters and calculate the penalized NPMLE corresponding to each $\gamma$. To search for the optimal tuning parameter $\gamma^*$, we choose the Bayesian information criterion (BIC) as our model selection criterion:
\begin{equation*}
    \text{BIC}(\gamma) = -2 \log \widehat{L}_{\mathrm{obs}}(\mathfrak{B}_\gamma) + \log(n) \vert \mathfrak{B}_\gamma \vert,
\end{equation*}
where $\widehat{L}_{\mathrm{obs}}(\mathfrak{B}_\gamma)$ is the maximum value of observed likelihood for the active set $\mathfrak{B}_\gamma$.

\section{Simulation studies}
\label{sec:simul}

\subsection{Unpenalized estimation}
In this subsection, we evaluate the empirical performance of the proposed unpenalized methods and two existing methods, namely complete-case analysis and single imputation. 

We set $p=4$ and generated $\boldsymbol{X}$ from the multivariate normal distribution with $\boldsymbol{\mu}=\boldsymbol{0}$ and $\boldsymbol{\Sigma}\equiv(0.5^{\vert i-j \vert})_{i,j=1,\ldots,p}$. We set $\boldsymbol{\beta} = (0.5, 0.5, 0.5, 0.5)^\mathrm{T}$ and $\Lambda(t) = 0.04t^{5/4}$. We set the censoring time to be $\min\{C^*,50\}$, where $C^*\sim\mathrm{Exp}(0.03)$; the censoring rate is approximately 34\%.

We considered a sample size of $n=500$ or 1000. For each subject, either all covariates are observed, or the first two covariates are missing. We considered two missing mechanisms, namely MCAR and MAR. For MCAR, subjects with missing data are randomly assigned. For MAR, we mimic a case-cohort study, where a subcohort consisting of 30\% of the whole sample was set to have observed covariates. Then, we randomly selected subjects outside the subcohort with $\Delta=1$ to have observed covariates. If all subjects with $\Delta=1$ were selected and the missing proportion is still higher than the desired level, then we randomly selected subjects with $\Delta=0$ to yield the desired missing proportion. We considered missing proportions ($p_M$) of 20\% and 40\%. 

We considered the proposed NPMLE, complete-case analysis under the standard Cox regression, and single imputation. For single imputation, we estimated $\boldsymbol{\mu}$ and $\boldsymbol{\Sigma}$ using MLE based on the fully observed $\boldsymbol{X}_i$'s, imputed the missing entries by their estimated conditional mean given the partially observed $\boldsymbol{X}_i$'s, and then fitted the standard Cox model on the imputed data. For the proposed method, we use bootstrap to obtain standard error estimators and confidence intervals for the regression parameters, with 500 bootstrap replicates. We considered 500 simulation replicates.
\begin{table}[!ht]
\renewcommand{\tabcolsep}{3bp}
\renewcommand{\arraystretch}{1.0}
\centering
\begin{threeparttable}
\begin{tabular}{lcrcccrcrc}
\toprule
  \multicolumn{2}{c}{}&\multicolumn{4}{c}{NPMLE} &\multicolumn{2}{c}{Complete Case} &\multicolumn{2}{c}{Single Imputation} \\
\cmidrule(lr){3-6}\cmidrule(lr){7-8}\cmidrule(lr){9-10}
Setting &Parameter &\multicolumn{1}{c}{Bias} &\multicolumn{1}{c}{SE} &\multicolumn{1}{c}{SEE} &\multicolumn{1}{c}{CP} &\multicolumn{1}{c}{Bias} &\multicolumn{1}{c}{SE} &\multicolumn{1}{c}{Bias} &\multicolumn{1}{c}{SE} \\
\midrule
$n=500$ &$\beta_1$ &$-$0.0022 &0.0771 &0.0764 &0.93 &$-$0.0017 &0.0782 &$-$0.0206 &0.0752\\
$p_M=20\%$ &$\beta_2$ &0.0059 &0.0856 &0.0854 &0.93 &0.0064 &0.0857 &$-$0.0130 &0.0816\\
&$\beta_3$ &0.0061 &0.0772 &0.0801 &0.94 &0.0051 &0.0824 &$-$0.0303 &0.0772\\
&$\beta_4$ &0.0051 &0.0701 &0.0724 &0.95 &0.0062 &0.0752 &$-$0.0241 &0.0698\\
\midrule
$n=500$ &$\beta_1$ &$-$0.0009 &0.0901 &0.0886 &0.93 &0.0014 &0.0914 &$-$0.0356 &0.0843\\
$p_M=40\%$ &$\beta_2$ &0.0061 &0.0965 &0.0995 &0.94 &0.0076 &0.0982 &$-$0.0290 &0.0896\\
&$\beta_3$ &0.0082 &0.0825 &0.0859 &0.95 &0.0086 &0.0949 &$-$0.0586 &0.0836\\
&$\beta_4$ &0.0049 &0.0773 &0.0776 &0.93 &0.0078 &0.0860 &$-$0.0486 &0.0750\\
\midrule
$n=1000$ &$\beta_1$ &0.0055 &0.0547 &0.0534 &0.94 &0.0057 &0.0553 &$-$0.0148 &0.0532\\
$p_M=20\%$ &$\beta_2$ &$-$0.0006 &0.0587 &0.0592 &0.95 &$-$0.0005 &0.0585 &$-$0.0210 &0.0551\\
&$\beta_3$ &0.0066 &0.0529 &0.0558 &0.95 &0.0075 &0.0580 &$-$0.0305 &0.0525\\
&$\beta_4$ &0.0007 &0.0514 &0.0502 &0.92 &0.0007 &0.0548 &$-$0.0287 &0.0513\\
\midrule
$n=1000$ &$\beta_1$ &0.0056 &0.0629 &0.0615 &0.93 &0.0062 &0.0636 &$-$0.0304 &0.0588\\
$p_M=40\%$ &$\beta_2$ &0.0000 &0.0689 &0.0681 &0.94 &0.0007 &0.0695 &$-$0.0361 &0.0633\\
&$\beta_3$ &0.0081 &0.0554 &0.0596 &0.95 &0.0092 &0.0660 &$-$0.0594 &0.0548\\
&$\beta_4$ &0.0005 &0.0547 &0.0536 &0.92 &0.0006 &0.0625 &$-$0.0530 &0.0528\\
\bottomrule
\end{tabular}
\begin{tablenotes}
\item Note: ``Bias'' is the empirical bias; ``SE'' is the empirical standard error; ``SEE'' is the average standard error estimate; ``CP'' is the empirical coverage probability of a 95\% confidence interval.
\end{tablenotes}
\caption{Results for unpenalized estimators of $\boldsymbol{\beta}$ under MCAR}
\label{table:unpenalizedMCAR}
\end{threeparttable}
\end{table}

The results for MCAR are shown in Table \ref{table:unpenalizedMCAR}. Both the NPMLE and complete-case analysis yield unbiased estimation, whereas single imputation yields noticeably larger bias than the other two methods, especially under a missing proportion of 40\%. This is because the imputed values are necessarily not as associated with the event time as the actual values, so the estimators are biased towards zero. Under all settings, single imputation yields the smallest standard error, followed by NPMLE. This is because single imputation trades some bias for efficiency, and the NPMLE uses more subjects than the complete-case analysis. For NPMLE, the standard errors of $\widehat{\beta}_3$ and $\widehat{\beta}_4$ tend to be smaller than those of $\widehat{\beta}_1$ and $\widehat{\beta}_2$, because $X_3$ and $X_4$ have more observations than $X_1$ and $X_2$ and thus have a larger ``effective sample size.''

\begin{figure}[!ht]
    \centering
    \includegraphics[width=16.5cm]{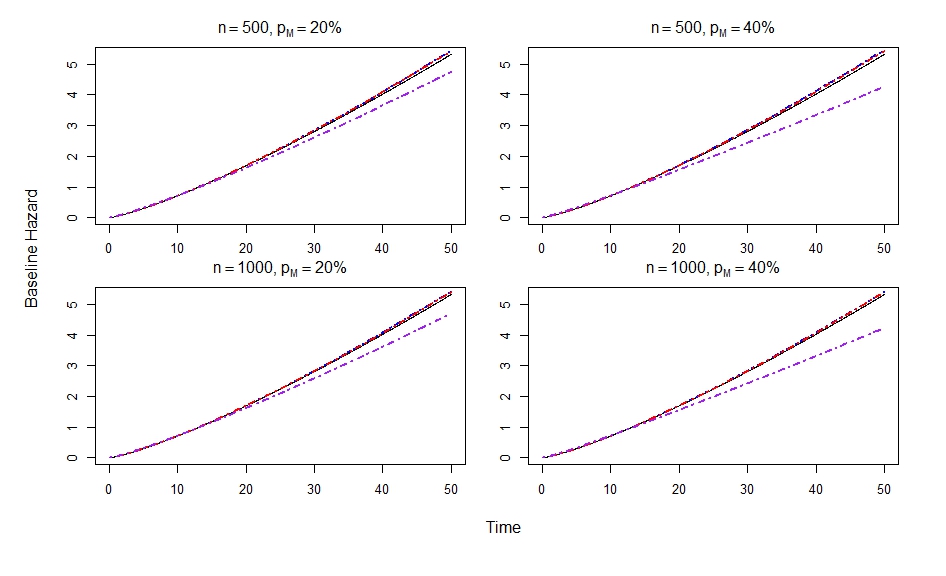}
    -------- True \quad\quad
    {\color{red}\tikz[baseline=-0.5ex]\draw [thick,dash dot] (0,0) -- (1
    ,0);} NPMLE \quad\quad
    {\color{blue}\tikz[baseline=-0.5ex]\draw [thick,dash dot] (0,0) -- (1
    ,0);} Complete Case \quad\quad
    {\color{mypurple}\tikz[baseline=-0.5ex]\draw [thick,dash dot] (0,0) -- (1
    ,0);} Single Imputation
    \caption{Results for unpenalized estimators of $\Lambda$ under MCAR.}
    \label{fig:unpenalizedMCAR}
\end{figure}

In Figure \ref{fig:unpenalizedMCAR}, we present the average value of $\Lambda$ over the replicates for different methods under MCAR. In this case, both the NPMLE and complete-case analysis are unbiased for $\Lambda$, whereas single imputation yields a biased estimator.

\begin{table}[!ht]
\renewcommand{\tabcolsep}{3bp}
\renewcommand{\arraystretch}{1.0}
\centering
\begin{threeparttable}
\begin{tabular}{lcrcccrcrc}
\toprule
  \multicolumn{2}{c}{}&\multicolumn{4}{c}{NPMLE} &\multicolumn{2}{c}{Complete Case} &\multicolumn{2}{c}{Single Imputation} \\
\cmidrule(lr){3-6}\cmidrule(lr){7-8}\cmidrule(lr){9-10}
Setting &Parameter &\multicolumn{1}{c}{Bias} &\multicolumn{1}{c}{SE} &\multicolumn{1}{c}{SEE} &\multicolumn{1}{c}{CP} &\multicolumn{1}{c}{Bias} &\multicolumn{1}{c}{SE} &\multicolumn{1}{c}{Bias} &\multicolumn{1}{c}{SE} \\
\midrule
$n=500$ &$\beta_1$ &$-$0.0012 &0.0773 &0.0751 &0.93 &$-$0.0524 &0.0709 &$-$0.0510 &0.0710\\
$p_M=20\%$ &$\beta_2$ &0.0070 &0.0848 &0.0839 &0.93 &$-$0.0455 &0.0778 &$-$0.0654 &0.0792\\
&$\beta_3$ &0.0046 &0.0761 &0.0791 &0.94 &$-$0.0486 &0.0753 &0.0181 &0.0770\\
&$\beta_4$ &0.0046 &0.0691 &0.0707 &0.94 &$-$0.0484 &0.0682 &$-$0.0157 &0.0719\\
\midrule
$n=500$ &$\beta_1$ &$-$0.0017 &0.0906 &0.0867 &0.92 &$-$0.0541 &0.0838 &$-$0.0678 &0.0794\\
$p_M=40\%$ &$\beta_2$ &0.0071 &0.0943 &0.0973 &0.94 &$-$0.0464 &0.0871 &$-$0.0877 &0.0865\\
&$\beta_3$ &0.0078 &0.0824 &0.0848 &0.94 &$-$0.0458 &0.0896 &-0.0079 &0.0837\\
&$\beta_4$ &0.0046 &0.0726 &0.0758 &0.95 &$-$0.0458 &0.0763 &$-$0.0412 &0.0735\\
\midrule
$n=1000$ &$\beta_1$ &0.0043 &0.0517 &0.0522 &0.94 &$-$0.0479 &0.0484 &$-$0.0467 &0.0490\\
$p_M=20\%$ &$\beta_2$ &$-$0.0027 &0.0591 &0.0582 &0.95 &$-$0.0550 &0.0532 &$-$0.0757 &0.0534\\
&$\beta_3$ &0.0076 &0.0523 &0.0549 &0.96 &$-$0.0450 &0.0504 &0.0215 &0.0550\\
&$\beta_4$ &0.0009 &0.0505 &0.0491 &0.92 &$-$0.0517 &0.0495 &$-$0.0195 &0.0520\\
\midrule
$n=1000$ &$\beta_1$ &0.0051 &0.0621 &0.0600 &0.92 &$-$0.0481 &0.0572 &$-$0.0629 &0.0549\\
$p_M=40\%$ &$\beta_2$ &$-$0.0016 &0.0702 &0.0670 &0.93 &$-$0.0559 &0.0636 &$-$0.0962 &0.0628\\
&$\beta_3$ &0.0074 &0.0569 &0.0587 &0.95 &$-$0.0476 &0.0603 &$-$0.0066 &0.0577\\
&$\beta_4$ &0.0009 &0.0523 &0.0524 &0.94 &$-$0.0511 &0.0568 &$-$0.0457 &0.0512\\
\bottomrule
\end{tabular}
\begin{tablenotes}
\item Note: See Note to Table \ref{table:unpenalizedMCAR}.
\end{tablenotes}
\caption{Results for unpenalized estimators of $\boldsymbol{\beta}$ under MAR}
\label{table:unpenalizedMAR}
\end{threeparttable}
\end{table}

The simulation results for MAR are shown in Table {\ref{table:unpenalizedMAR}}. Under MAR, NPMLE is unbiased, whereas the complete-case analysis and single imputation are biased. Similar to the MCAR setting, single imputation yields the smallest standard errors overall.  Under NPMLE, the standard errors of $\widehat{\beta}_3$ and $\widehat{\beta}_4$ tend to be smaller than those of $\widehat{\beta}_1$ and $\widehat{\beta}_2$. Note that NPMLE yields the smallest mean squared error among all three methods in all settings, under MCAR and MAR.

\begin{figure}[!ht]
    \centering
    \includegraphics[width=16.5cm]{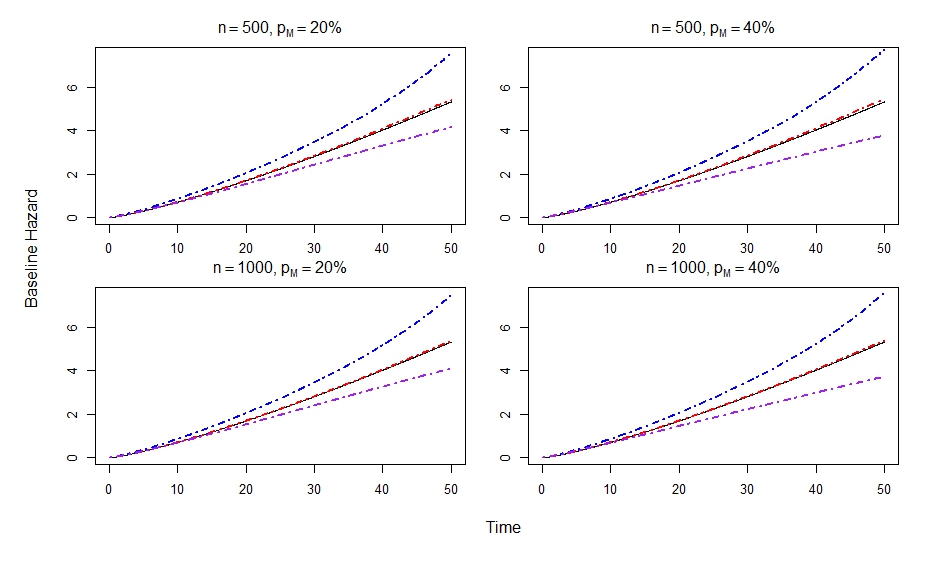}
    -------- True \quad\quad
    {\color{red}\tikz[baseline=-0.5ex]\draw [thick,dash dot] (0,0) -- (1
    ,0);} NPMLE \quad\quad
    {\color{blue}\tikz[baseline=-0.5ex]\draw [thick,dash dot] (0,0) -- (1
    ,0);} Complete Case \quad\quad
    {\color{mypurple}\tikz[baseline=-0.5ex]\draw [thick,dash dot] (0,0) -- (1
    ,0);} Single Imputation
    \caption{Results for unpenalized estimators of $\Lambda$ under MAR.}
    \label{fig:unpenalizedMAR}
\end{figure}

In Figure \ref{fig:unpenalizedMAR}, we present the average value of $\Lambda$ over the replicates for different methods under MAR. The NPMLE yields unbiased estimation, whereas both the complete-case analysis and single imputation are biased.

\subsection{Penalized estimation}
In this subsection, we compare the performance of penalized methods. We considered a sample size of $n=500$ or $1000$ and a number of covariates of $p=100$. We draw the covariates from the multivariate normal distribution with $\boldsymbol{\mu} = \boldsymbol{0}$ and $\boldsymbol{\Sigma} = \text{diag}(\boldsymbol{\Sigma}_1, \boldsymbol{\Sigma}_2)$, where $\boldsymbol{\Sigma}_1=(0.2^{\vert i-j\vert})_{i,j=1,\ldots,50}$ and $\boldsymbol{\Sigma}_2=(0.5^{\vert i-j\vert})_{i,j=1,\ldots,50}$. For the survival model, we set $\boldsymbol{\beta} = (\underbrace{0.25, \ldots, 0.25}_{4},\underbrace{0, \ldots, 0}_{192}, \underbrace{0.25, \ldots, 0.25}_{4})^\mathrm{T}$ and $\Lambda(t) = 0.04t^{5/4}$. We set the censoring time to be $\min\{C^*,50\}$, where $C^*\sim\mathrm{Exp}(0.035)$; the censoring rate is approximately 34\%. For each subject, either all covariates are observed, or only the covariates with even indices (i.e., $X_2,X_4,\ldots,X_{100}$) are observed. We considered missing mechanisms of MCAR and MAR, generated in the same way as the unpenalized case.

We considered the penalized NPMLE, the complete-case analysis, and single imputation. For single imputation, we imputed the missing values in the same way as for the unpenalized case. For the complete-case analysis and single imputation, we obtained the active sets using maximum penalized partial likelihood estimation with a LASSO penalty using the observed or completed data and then refitted the model over the active sets. BIC was used in choosing the best model for all three methods.  We report the true positive rate (TPR), false negative rate (FDR), and mean squared error (MSE) for each method. These statistics, based on 500 simulation replicates, are summarized in Table \ref{table:penalized}.
The results, based on 500 simulation replicates, are summarized in Table \ref{table:penalized}.

\begin{table}[!ht]
\renewcommand{\tabcolsep}{3bp}
\renewcommand{\arraystretch}{1.0}
\centering
\begin{threeparttable}
\begin{tabular}{lcccccccc}
\toprule
  \multicolumn{3}{c}{}&\multicolumn{3}{c}{$p_M=20\%$} &\multicolumn{3}{c}{$p_M=40\%$} \\
\cmidrule(lr){4-6}\cmidrule(lr){7-9}
 \multicolumn{1}{c}{$n$} &Pattern &Method  &TPR &FDR &MSE  &TPR &FDR &MSE\\
\midrule
500 &MCAR &Penalized NPMLE &0.9335 &0.1012 &0.1013 &0.8647 &0.1233 &0.1517\\
& &Complete Case  &0.9160 &0.1012 &0.1219 &0.8472 &0.1231 &0.1934\\
& &Single Imputation &0.9322 &0.0917 &0.0978 &0.8502 &0.0889 &0.1420\\
\midrule
500 &MAR &Penalized NPMLE &0.9165 &0.1092 &0.1126 &0.8417 &0.1283 &0.1680\\
& &Complete Case &0.8985 &0.1092 &0.1126 &0.8112 &0.1283 &0.1785\\
& &Single Imputation &0.8867 &0.0810 &0.1207 &0.7800 &0.0902 &0.1805\\
 \midrule
1000 &MCAR &Penalized NPMLE &0.9948 &0.0814 &0.0367 &0.9820 &0.0921 &0.0498\\
& &Complete Case &0.9903 &0.0810 &0.0432 &0.9738 &0.0917 &0.0677\\
& &Single Imputation &0.9948 &0.0760 &0.0352 &0.9785 &0.0681 &0.0472\\
\midrule
1000 &MAR &Penalized NPMLE &0.9918 &0.0770 &0.0373 &0.9725 &0.0926 &0.0532\\
& &Complete Case &0.9845 &0.0768 &0.0439 &0.9605 &0.0926 &0.0672\\
& &Single Imputation &0.9793 &0.0538 &0.0452 &0.9355 &0.0654 &0.0730\\
\bottomrule
\end{tabular}
\begin{tablenotes}
\item Note: ``TPR'' is the true positive rate; ``FDR'' is the false negative rate; ``MSE'' is the mean squared error.
\end{tablenotes}
\caption{Results for penalized estimators of $\boldsymbol{\beta}$ under MCAR and MAR}
\label{table:penalized}
\end{threeparttable}
\end{table}

The penalized NPMLE has the highest TPR and FDR among all three methods. This implies that the penalized NPMLE tends to select more covariates, both relevant and irrelevant ones. In term of MSE, penalized NPMLE is uniformly better than the complete-case analysis, especially for MCAR. Single imputation could yield smaller MSE than penalized NPMLE under MCAR, probably because single imputation is biased towards zero, as demonstrated in the simulation studies for the unpenalized case. By contrast, single imputation always yields higher MSE than the penalized NPMLE under MAR.

\subsection{Unpenalized and penalized estimation under a misspecified covariate distribution}
To evaluate the sensitivity of the proposed methods to the normality assumption, we conduct simulation studies with a misspecified covariate distribution. In particular, we generated the multivariate normal random vector as described above. Then, we transformed each component of the normal vector by $F_5^{-1}\circ\Phi$ and set the transformed value as a covariate, where $\Phi$ and $F_5$ are the cumulative distribution functions of the standard normal distribution and the $t$ distribution with five degrees of freedom, respectively. As a result, each covariate marginally follows a $t$ distribution. The event and censoring times were then generated in the same way as the above. We considered both the unpenalized and penalized estimators. The results are shown in Tables 4, 5, and 6. 

\begin{table}[!ht]
\renewcommand{\tabcolsep}{3bp}
\renewcommand{\arraystretch}{1.0}
\centering
\begin{threeparttable}
\begin{tabular}{lcrcccrcrc}
\toprule
  \multicolumn{2}{c}{}&\multicolumn{4}{c}{NPMLE} &\multicolumn{2}{c}{Complete Case} &\multicolumn{2}{c}{Single Imputation} \\
\cmidrule(lr){3-6}\cmidrule(lr){7-8}\cmidrule(lr){9-10}
Setting &Parameter &\multicolumn{1}{c}{Bias} &\multicolumn{1}{c}{SE} &\multicolumn{1}{c}{SEE} &\multicolumn{1}{c}{CP} &\multicolumn{1}{c}{Bias} &\multicolumn{1}{c}{SE} &\multicolumn{1}{c}{Bias} &\multicolumn{1}{c}{SE} \\
\midrule
$n=500$ &$\beta_1$ &$-$0.0020 &0.0637 &0.0632 &0.93 &$-$0.0005 &0.0648 &$-$0.0266 &0.0615\\
$p_M=20\%$ &$\beta_2$ &0.0032 &0.0710 &0.0703 &0.94 &0.0053 &0.0711 &$-$0.0221 &0.0668\\
&$\beta_3$ &0.0070 &0.0653 &0.0669 &0.95 &0.0051 &0.0686 &$-$0.0456 &0.0695\\
&$\beta_4$ &0.0072 &0.0595 &0.0605 &0.94 &0.0072 &0.0636 &$-$0.0323 &0.0596\\
\midrule
$n=500$ &$\beta_1$ &$-$0.0026 &0.0745 &0.0732 &0.94 &0.0023 &0.0765 &$-$0.0464 &0.0681\\
$p_M=40\%$ &$\beta_2$ &0.0014 &0.0780 &0.0817 &0.95 &0.0066 &0.0789 &$-$0.0432 &0.0704\\
&$\beta_3$ &0.0107 &0.0710 &0.0727 &0.94 &0.0087 &0.0798 &$-$0.0815 &0.0777\\
&$\beta_4$ &0.0077 &0.0657 &0.0658 &0.92 &0.0094 &0.0732 &$-$0.0624 &0.0640\\
\midrule
$n=1000$ &$\beta_1$ &0.0041 &0.0448 &0.0439 &0.92 &0.0057 &0.0454 &$-$0.0225 &0.0429\\
$p_M=20\%$ &$\beta_2$ &$-$0.0004 &0.0487 &0.0483 &0.95 &0.0012 &0.0487 &$-$0.0269 &0.0447\\
&$\beta_3$ &0.0054 &0.0452 &0.0461 &0.94 &0.0046 &0.0489 &$-$0.0485 &0.0488\\
&$\beta_4$  &0.0019 &0.0418 &0.0417 &0.94 &0.0012 &0.0443 &$-$0.0381 &0.0452\\
\midrule
$n=1000$ &$\beta_1$ &0.0030 &0.0513 &0.0505 &0.93 &0.0068 &0.0524 &$-$0.0424 &0.0467\\
$p_M=40\%$ &$\beta_2$ &$-$0.0023 &0.0557 &0.0554 &0.93 &0.0020 &0.0568 &$-$0.0476 &0.0501\\
&$\beta_3$ &0.0082 &0.0479 &0.0501 &0.93 &0.0059 &0.0551 &$-$0.0863 &0.0520\\
&$\beta_4$ &0.0026 &0.0455 &0.0453 &0.94 &0.0019 &0.0507 &$-$0.0685 &0.0483\\
\bottomrule
\end{tabular}
\begin{tablenotes}
\item Note: See Note to Table 1.
\end{tablenotes}
\caption{Results for unpenalized estimators of $\boldsymbol{\beta}$ with a misspecified distribution under MCAR}
\label{table:misspecUnpenalizedMCAR}
\end{threeparttable}
\end{table}

\begin{table}[!ht]
\renewcommand{\tabcolsep}{3bp}
\renewcommand{\arraystretch}{1.0}
\centering
\begin{threeparttable}
\begin{tabular}{lcrcccrcrc}
\toprule
  \multicolumn{2}{c}{}&\multicolumn{4}{c}{NPMLE} &\multicolumn{2}{c}{Complete Case} &\multicolumn{2}{c}{Single Imputation} \\
\cmidrule(lr){3-6}\cmidrule(lr){7-8}\cmidrule(lr){9-10}
Setting &Parameter &\multicolumn{1}{c}{Bias} &\multicolumn{1}{c}{SE} &\multicolumn{1}{c}{SEE} &\multicolumn{1}{c}{CP} &\multicolumn{1}{c}{Bias} &\multicolumn{1}{c}{SE} &\multicolumn{1}{c}{Bias} &\multicolumn{1}{c}{SE} \\
\midrule
$n=500$ &$\beta_1$ &$-$0.0011 &0.0625 &0.0610 &0.93 &$-$0.0445 &0.0587 &$-$0.0480 &0.0581\\
$p_M=20\%$ &$\beta_2$ &0.0083 &0.0681 &0.0680 &0.94 &$-$0.0371 &0.0636 &$-$0.0590 &0.0642\\
&$\beta_3$ &0.0044 &0.0643 &0.0652 &0.94 &$-$0.0395 &0.0632 &0.0027 &0.0684\\
&$\beta_4$ &0.0052 &0.0564 &0.0583 &0.94 &$-$0.0385 &0.0549 &$-$0.0219 &0.0629\\
\midrule
$n=500$ &$\beta_1$ &0.0002 &0.0759 &0.0703 &0.92 &$-$0.0452 &0.0711 &$-$0.0690 &0.0664\\
$p_M=40\%$ &$\beta_2$ &0.0050 &0.0773 &0.0783 &0.94 &$-$0.0422 &0.0733 &$-$0.0935 &0.0717\\
&$\beta_3$ &0.0090 &0.0680 &0.0706 &0.94 &$-$0.0386 &0.0714 &$-$0.0327 &0.0741\\
&$\beta_4$ &0.0048 &0.0610 &0.0634 &0.94 &$-$0.0428 &0.0655 &$-$0.0543 &0.0657\\
\midrule
$n=1000$ &$\beta_1$ &0.0047 &0.0423 &0.0421 &0.93 &$-$0.0394 &0.0403 &$-$0.0431 &0.0403\\
$p_M=20\%$ &$\beta_2$ &0.0009 &0.0481 &0.0467 &0.95 &$-$0.0439 &0.0445 &$-$0.0663 &0.0442\\
&$\beta_3$ &0.0049 &0.0432 &0.0448 &0.94 &$-$0.0382 &0.0422 &0.0039 &0.0481\\
&$\beta_4$  &0.0004 &0.0407 &0.0404 &0.93 &$-$0.0425 &0.0411 &$-$0.0283 &0.0435\\
\midrule
$n=1000$ &$\beta_1$ &0.0044 &0.0497 &0.0483 &0.93 &$-$0.0426 &0.0465 &$-$0.0667 &0.0445\\
$p_M=40\%$ &$\beta_2$ &$-$0.0011 &0.0561 &0.0536 &0.94 &$-$0.0489 &0.0518 &$-$0.0995 &0.0502\\
&$\beta_3$ &0.0065 &0.0464 &0.0484 &0.95 &$-$0.0411 &0.0494 &$-$0.0338 &0.0530\\
&$\beta_4$ &0.0007 &0.0427 &0.0437 &0.95 &$-$0.0459 &0.0451 &$-$0.0607 &0.0456\\
\bottomrule
\end{tabular}
\begin{tablenotes}
\item Note: See Note to Table 1.
\end{tablenotes}
\caption{Results for unpenalized estimators of $\boldsymbol{\beta}$ with a misspecified distribution under MAR}
\label{table:misspecUnpenalizedMAR}
\end{threeparttable}
\end{table}

\begin{table}[!ht]
\renewcommand{\tabcolsep}{3bp}
\renewcommand{\arraystretch}{1.0}
\centering
\begin{threeparttable}
\begin{tabular}{lcccccccc}
\toprule
  \multicolumn{3}{c}{}&\multicolumn{3}{c}{$p_M=20\%$} &\multicolumn{3}{c}{$p_M=40\%$} \\
\cmidrule(lr){4-6}\cmidrule(lr){7-9}
 \multicolumn{1}{c}{$n$} &Pattern &Method  &TPR &FDR &MSE  &TPR &FDR &MSE\\
\midrule
500 &MCAR &Penalized NPMLE &0.9848 &0.1034 &0.0535 &0.9465 &0.1238 &0.0823\\
& &Complete Case &0.9773 &0.1034 &0.0645 &0.9430 &0.1238 &0.1094\\
& &Single Imputation &0.9830 &0.0877 &0.0503 &0.9300 &0.0953 &0.0833\\
\midrule
500 &MAR &Penalized NPMLE &0.9790 &0.1077 &0.0558 &0.9285 &0.1285 &0.0939\\
& &Complete Case &0.9715 &0.1075 &0.0582 &0.9255 &0.1282 &0.0967\\
& &Single Imputation &0.9603 &0.0841 &0.0660 &0.8692 &0.0943 &0.1175\\
 \midrule
1000 &MCAR &Penalized NPMLE &0.9988 &0.0756 &0.0214 &0.9965 &0.0958 &0.0267\\
& &Complete Case &0.9988 &0.0752 &0.0242 &0.9948 &0.0954 &0.0379\\
& &Single Imputation &0.9988 &0.0675 &0.0208 &0.9960 &0.0635 &0.0265\\
\midrule
1000 &MAR &Penalized NPMLE &0.9985 &0.0852 &0.0216 &0.9953 &0.0976 &0.0281\\
& &Complete Case &0.9988 &0.0848 &0.0254 &0.9935 &0.0972 &0.0362\\
& &Single Imputation &0.9950 &0.0670 &0.0274 &0.9760 &0.0739 &0.0443\\
\bottomrule
\end{tabular}
\begin{tablenotes}
\item Note: See Note to Table 3.
\end{tablenotes}
\caption{Results for penalized estimators of $\boldsymbol{\beta}$ with a misspecified distribution under MCAR and MAR}
\label{table:misspecPenalized}
\end{threeparttable}
\end{table}

For the unpenalized methods, the results are similar in pattern as those under a correctly-specified covariate distribution. In particular, NPMLE and the complete-case analysis are unbiased under MCAR, while single imputation tends to be biased.  However, the standard errors under NPMLE and single imputation are of similar level, and single imputation does not have noticeable smaller standard errors than the other methods. In term of MSE, NPMLE dominates the other two methods under all settings.

For the penalized methods, similar to the results in Table \ref{table:penalized}, the penalized NPMLE has overall the highest TPR and FDR. Penalized NPMLE always has a smaller MSE than the complete-case analysis. Under MCAR, penalized NPMLE and single imputation have similar values of MSE. Under MAR, single imputation has the largest MSE among all three methods.

\section{Real data analysis}
We analyzed a dataset of kidney renal clear cell carcinoma (KIRC) from TCGA. The dataset, released in November 2015, was downloaded through the RTCGA package (\citeauthor{kosinski2016rtcga}, \citeyear{kosinski2016rtcga}) in R. In the study, times to new tumor events and death were collected, which were potentially subject to right censoring. Also, omic variables including gene expressions, measured by RNA sequencing, and protein expressions, measured by reverse-phase protein array, were collected for some or most subjects. In this paper, we focus on the association between time to death since initial diagnosis and the omic variables.

The dataset contains 20531 gene expressions and 217 protein expressions. There are 530 subjects with both survival data and gene expression measurements. Among these subjects, 475 have measurements in protein expressions. Following \citeauthor{zhao2015combining} (\citeyear{zhao2015combining}), we filtered out gene expressions with 0 median absolute deviation, resulting in the removal of 2865 genes. Then, following \cite{cancer2013comprehensive}, we selected the top 1500 gene expressions with the largest maximum absolute deviation. We then performed the $\log(1+x)$-transformation on the gene expressions. In addition, we removed 5 protein expressions that were missing for over 90\%  of the subjects. After the above preliminary processing, we performed supervised screening by fitting a separate Cox model for time to death against each gene or protein expression and selected the top 150 covariates with the smallest $p$-values; here, complete-case analysis was used in the presence of missing values. This resulted in 16 protein expressions and 134 gene expressions selected for downstream analyses. The missing proportion for protein expressions is 10\%, and the censoring rate is 58\%. 

We performed the penalized NPMLE, complete-case analysis, and single imputation approaches on the processed data. Note that we obtained the active set based on standardized covariates and refitted our models on the original unstandardized covariates. The analysis results are presented in Table 7. The penalized NPMLE, complete-case analysis, and single imputation selected 8, 4, and 5 covariates, respectively. This is consistent with the findings in the simulation studies that the penalized NPMLE tends to select the largest number of features.

\begin{table}[!ht]
\renewcommand{\tabcolsep}{3bp}
\renewcommand{\arraystretch}{1.0}
\centering
\begin{threeparttable}
\begin{tabular}{lrrr}
\toprule
  Variable &\multicolumn{1}{r}{Penalized NPMLE} &\multicolumn{1}{r}{Complete Case} &\multicolumn{1}{r}{Single Imputation} \\
\midrule
Protein -- MAPK\_pT202\_Y204 &$-$0.3288 &$-$0.3209 &$-$0.3266\\
Gene -- CDCA3 (83461) &0.0797 &0.1487 &0.1172\\
Gene -- SHOX2 (6474) &0.1089 &0.1608 &0.1432\\
Gene -- LOC286467 (286467) &0.0883 &0.1642 &0.1338\\
Gene -- DNASE1L3 (1776) &$-$0.0976 &$\boldsymbol{\cdot}$ &$-$0.1261\\
Gene -- BRD9 (65980) &0.1717 &$\boldsymbol{\cdot}$ &$\boldsymbol{\cdot}$\\
Gene -- PHF21A (51317) &0.4149 &$\boldsymbol{\cdot}$ &$\boldsymbol{\cdot}$\\
Gene -- CARS (833) &0.1477 &$\boldsymbol{\cdot}$ &$\boldsymbol{\cdot}$\\
\bottomrule
\end{tabular}
\begin{tablenotes}
\item Note: For gene expressions, the Entrez IDs are given in the parentheses.
\end{tablenotes}
\caption{Regression parameter estimates for the KIRC data}
\label{table:realDataResult}
\end{threeparttable}
\end{table}

We evaluate the prediction performance of the three methods as follows. 
We randomly split the data into training and testing sets with a 7$:$3 ratio of sample sizes and performed the three estimation procedures on the training data.
Then, to facilitate evaluation of the fitted models, we imputed the missing values in the testing data by single imputation, where the whole data set was used to estimate the imputation model. We calculated the concordance index (C-index) (\citeauthor{harrell1982evaluating}, \citeyear{harrell1982evaluating}) between the event time and the estimated $\boldsymbol{X}^{\mathrm{T}}\boldsymbol{\beta}$ on the (imputed) testing data. The above procedure was repeated 100 times. Note that we imputed the testing data in the exact same way as in the single imputation method in the simulation studies. The average C-index values over the 100 splits for penalized NPMLE, complete-case analysis, and single imputation are 0.660, 0.656, and 0.658, respectively. The C-index values are similar due to the small missing proportion, with the proposed method having a slight advantage.

\section{Discussion}
In this paper, we propose a likelihood-based approach for (penalized) estimation of the Cox proportional hazards model, where covariates could be missing. We devise a novel EM algorithm that enables efficient computation under arbitrary missing patterns and a large number of missing covariates. Instead of performing multi-dimensional numerical integration over all dimensions of the missing covariates, we propose a linear transformation on the covariates, so that the expectations of all but one components of the transformed variables have closed-form expressions.

As for likelihood-based methods in general, we need to impose modeling assumptions on the missing covariates (except when only a few covariates are involved, in which case a full nonparametric model can be fit). The proposed methods depend crucially on the Gaussian assumptions on the covariates; without these assumptions, the transformation approach to reduce the dimension of numerical integration is not applicable. One possible approach to relax the Gaussian assumptions is to assume that the observed covariates are transformed values of underlying Gaussian variables, that is, $X_j=g_j(Z_j)$ for some transformation function $g_j$ and Gaussian variable $Z_j$. We then assume that the dependence between the outcome and covariates is mediated through $\boldsymbol{Z}\equiv(Z_1,\ldots,Z_p)^{\mathrm{T}}$, such that $\lambda(t\mid\boldsymbol{X},\boldsymbol{Z})=\lambda(t)e^{\boldsymbol{\beta}^{\mathrm{T}}\boldsymbol{Z}}$.
In this scenario, the proposed transformation approach can still be adopted.

The proposed transformation technique can also be applied to random effect models with Gaussian latent variables (\citeauthor{papageorgiou2019overview}, \citeyear{papageorgiou2019overview}; \citeauthor{sun2019regularized}, \citeyear{sun2019regularized}; \citeauthor{wong2022semiparametric}, \citeyear{wong2022semiparametric}). In general, we can accommodate an outcome variable that follows a survival model or a generalized linear model that regresses on a linear combination of Gaussian latent variables. This outcome variable can also be jointly modelled with other Gaussian outcomes that regress linearly on the random effects. To compute the MLE, we can develop a similar EM algorithm, where in the E-step, we transform the latent variable vector such that the first component is the linear combination present in the survival or generalized linear model.

Due to its flexibility, multiple imputation is a popular approach for handling missing data. The proposed methods have advantages over multiple imputation approaches, especially when variable selection is desirable, in two key respects.  First, under multiple imputation, it is often difficult to explicitly define the estimator, as it is typically the limit of some iterative algorithm. This makes theoretical studies of the estimator very challenging. By contrast, the proposed estimator is the maximizer of the (penalized) likelihood, and existing techniques (such as \citeauthor{wang2007unified}, \citeyear{wang2007unified}) can be applied to establish the theoretical properties. Second, multiple imputation is not amenable to simultaneous variable selection and estimation. One advantage of penalization methods is that they perform variable selection and estimation simultaneously: penalization shrinks the estimators towards zero, and some estimators are shrunk to exactly zero, thereby eliminating the corresponding covariates. However, even though penalized estimation can be performed on each imputed dataset to yield a sparse estimator, the final estimator that combines results from all imputed datasets is generally not sparse, as a variable would be selected even if it is selected in just one of the imputed dataset. By contrast, because the proposed method imposes a penalty on a single likelihood, it performs simultaneous variable selection and estimation.

In the proposed methods, we fit an unstructured covariance matrix for the covariates and estimate it by unpenalized MLE. As a result, we cannot accommodate a high-dimensional setting with $p>n$, as the variance estimator would be non-positive definite. To accommodate high-dimensional data, one can consider shrinkage estimators for covariance estimation (\citeauthor{ledoit2004well}, \citeyear{ledoit2004well}; \citeauthor{warton2008penalized}, \citeyear{warton2008penalized}). Alternatively, we could impose structures on the covariance matrix to facilitate estimation. For example, we may fit a factor model for $\boldsymbol{X}$, such that $\boldsymbol{\Sigma}$ can be decomposed into a low-rank matrix plus a sparse or diagonal matrix (\citeauthor{fan2008high}, \citeyear{fan2008high}). These approaches would require modifications to the M-step of the proposed algorithm, but the E-step remains the same.


\bibliographystyle{apalike}

\end{document}